\documentstyle[12pt]{article}
\def\D{\Delta}
\def\tt{\tilde{t}}
\def\adt{\tilde{a}_d}

\def\G{\Gamma}
\def\r{\rho}
\def\f{\phi}
\def\s{\sigma}
\def\ydot{\dot{y}}
\def\t{\tau}
\def\ve{\varepsilon}
\def\d{\delta}
\def\p{\psi}
\def\o{\omega}
\def\m{\mu}
\def\n{\nu}
\newcommand{\calL}{{\cal L}}
\newcommand{\calD}{{\cal D}}
\newcommand{\bc}{\begin{centerline}}
\newcommand{\ec}{\end{centerline}}
\newcommand{\be}{\begin{equation}}
\newcommand{\ee}{\end{equation}}
\newcommand{\bea}{\begin{eqnarray}}
\newcommand{\eea}{\end{eqnarray}}
\newcommand{\nen}{\nonumber \\ \relax}

\newcommand{\forcepar}{{\hskip 10pt\vskip -15pt}}
\newfont{\headfont}{cmbx10 scaled 1440}
\newfont{\headfontb}{cmbx10 scaled 1200}
\newfont{\namefont}{cmr9}
\newfont{\initialfont}{cmr9 scaled 1200}
\newfont{\addfont}{cmti10}

\newcommand{\inv}[1]{\frac{1}{#1}}

\newcommand{\IR}{{I \kern -0.4em R}}
\newcommand{\IC}{{I \kern -0.65em C}}

\newcommand{\prd}[1]{{\it Phys. Rev.} {\bf D#1}}
\newcommand{\prb}[1]{{\it Phys. Rev.} {\bf B#1}}
\newcommand{\ap}[1]{{\it Ann. Phys.} {\bf #1}}

\newcommand{\cqg}[1]{{\it Class. Quantum Grav.} {\bf #1}}

\newcommand{\pa}[1]{{\it Physica} {\bf A#1}}
\newcommand{\pto}[1]{{\it Phys. Today} {\bf #1}}
\newcommand{\prep}[1]{{\it Phys. Rep.} {\bf #1}}
\begin{document}

\begin{titlepage}
\renewcommand{\thefootnote}{\fnsymbol{footnote}}
\begin{center}
{\headfont Fluctuations of the Unruh Temperature}
\footnote{This work is supported in part by funds provided by the
U. S. Department of Energy (D.O.E.) under cooperative agreement
\#DE-FC02-94ER40818.}
\end{center}
\vskip 0.3truein
\begin{center}
{ {\initialfont J}{\namefont EAN-}{\initialfont G}{\namefont UY}
		    {\initialfont D}{\namefont EMERS  } }
\end{center}
\begin{center}
{\addfont{Center for Theoretical Physics,}}\\
{\addfont{Laboratory for Nuclear Science}}\\
{\addfont{and Department of Physics,}}\\
{\addfont{Massachusetts Institute of Technology}}\\
{\addfont{Cambridge, Massachusetts 02139 U.S.A.}}\\
{\addfont{email: jgdemers@mitlns.mit.edu}}
\end{center}
\vskip 0.5truein
\begin{abstract}

Using the influence functional formalism, the problem of an
accelerating
detector in the presence of
a scalar field in its ground state is considered in Minkowski space.
As is known since the work of Unruh,  to a quantum mechanical
detector following a definite,   classical acceleration,
the field appears to be thermally excited.
We relax the requirement of perfect classicality for the trajectory
and substitute it with
 one of {\it derived} classicality through
the criteria of decoherence.
The ensuing fluctuations in
temperature are then related with   the time and the amplitude
of excitation in the
detector's internal degree of freedom.

\vskip 0.5truein
\leftline {MIT-CTP  2342  \hfill August 1994}
\smallskip
\end{abstract}

\end{titlepage}
\setcounter{footnote}{0}
\section{}\label{intro}
\forcepar
The influence  functional (IF) formalism was developed 30 years
ago by Feynman and Vernon\cite{Fey1} to study the properties
of quantum open systems. It has since been usefully applied
to a number of physical problems (including masers, polaron,
transport, localization) and was also shown to be closely
related with the close-time-path formalism\cite{ZCYC}. More recently,
it was used by quantum cosmologists to study the conditions
under which a quantum mechanical system decoheres and can be
said to behave classically owing to its  interaction with the
unobserved environment\cite{Zu}.
The use of the IF in that context  can be seen  in the evolution
of the reduced density matrix.

Suppose the system and the environment have each a single degree
of freedom, that they  are described by the actions
$S_{sys}[x]$ and $S_{env}[y]$ respectively and that their interaction
is described by the action $S_{int}[x,y]$.
Suppose also that at $t=0$, the system and the environment are
not correlated so that the total density matrix is separable:
\be
\r(x_1,y_1,x_2,y_2;t=0)=\r_{sys}(x_1,x_2;t=0)\times \r_{env}(y_1,y_2;t=0).
\ee
Then at time $t$ later, the reduced density matrix (where the
environment is integrated out) will be  given by\cite{Hal,Hu1}:
\be
\r_{red}(x_1^f,x_2^f;t)=\int dx_1^i\:\int dx_2^i\:
P(x_1^f,x_2^f,t|x_1^i,x_2^i,0)\:\r_{sys}(x_1^i,x_2^i;0),
\ee
with the kernel
\be
P(x_1^f,x_2^f,t|x_1^i,x_2^i,0)=\int_{x_1^i}^{x_1^f}
\calD x_1\:\int_{x_2^i}^{x_2^f} \calD x_2  \:e^{i(S_{sys}[x_1]-
S_{sys}[x_2])}\:\:e^{i\G[x_1,x_2]},
\label{p}
\ee
where   the  boundary conditions on the functional integrals
are given for  times $0$ and $t$ (our units are such
that $\hbar=c=1$).
The IF is the last term in (\ref{p}). It is  a property of
the environment and the the way it is coupled to the system
and its formal expression is
\bea
e^{i\G[x_1,x_2]}&=&\int dy^f  dy_1^i dy^i_2\:\int_{y_1^i}^{y^f}
\calD y_1\:\int_{y_2^i}^{y^f} \calD y_2\:\:e^{i(S_{env}[y_1]+
S_{int}[x_1,y_1])}\nen
&{}&\hspace{0.5 in}\times\   e^{-i(S_{env}[y_2] +
S_{int}[x_2,y_2])}\:\r_{env}(y_1^i,y_2^i;0).
\label{if}
\eea
If at $t=0$, the environment is in a pure state then (\ref{if})
takes the conceptually simple form
\be
e^{i\G[x_1,x_2]}=\langle\p_2(t)|\p_1(t)\rangle.
\label{if2}
\ee
Here, $|\p_1(t)\rangle$ is the state that   evolved from the
initial state at $t=0$ under the dynamics dictated  by
$S_{env}[y_1]+S_{int}[x_1,y_1]$ in which  $x_1$ acts  as a c-number,
time dependant source; likewise   $|\p_2(t)\rangle$ is
governed by $S_{env}[y_2]+S_{int}[x_2,y_2]$.
For an environment made of many non-interacting and initially
uncorrelated particles, it is easy to see that the total $\G$
is simply the sum of the contributions from each particle.
Unless the paths $[x_1(t)]$ and  $[x_2(t)]$ are such that the
states in (\ref{if2}) are adiabaticlly disturbed, $\G$ will
have not only a real part  $\G_R$ but also a (positive)
imaginary part   $\G_I$.

In the next section, we review how the IF predicts the heating
of the  scalar field forming the  environment of a detector
on a uniformly accelerating course\cite{Un,Anglin}.  This is
achieved by assuming that while the detector's internal degree
of freedom is quantum mechanical, its position degree of
freedom remains strictly   classical.
In Section~\ref{fluc}, we consider  a different limit.
The spacetime   trajectory is  taken to be a {\it decohered}
spacetime path, with a spread around the uniformly accelerating
trajectory. Focusing on this position degree of freedom
and taking now the detector's excitation to be fixed, we  obtain
a relation for the spread in acceleration involved in
constructing the approximately classical path.

As usual,  environment-induced  decoherence is indicated by the
diagonalization of the reduced density matrix. Thus,
in the position basis, decoherence at a time t is  a measure of how
$\r_{red}(x_1^f,x_2^f;t)$ has become diagonal. It is clear that
if  $\G_I$ is large for   pair of paths $(x_1(t), x_2(t))$ that
are far apart from one another in spacetime, then the contribution
of these pairs   will be suppressed in (\ref{p}). As a result,
$\r_{red}(x_1^f,x_2^f;t)$ will  be more diagonal.  Our  study of
decoherence will thus be through $\G_I$.  Note that a decoherence
functional is sometimes defined to include coarse graining of the
system paths in addition to the tracing out of the
environment\cite{Hal,Har}. But in that case, diagonalization
of the IF defined here is still the  mechanism to achieve
decoherence.
In the last section, we  remark  on our result and
comment on their  relevance in the context of
black holes.

\vskip .5in
\section{}
\label{Unruh}
\forcepar
Consider an  environment made of a massless scalar field in
two spacetime dimensions\cite{Hu1,Anglin}. In a finite box
of   spatial dimension $L$, the mode expansion reads
\be
\f(x,t)=\sqrt{2\over L} \sum_k[y_k^+cos\: kx+y_k^-sin\: kx],
\ee
with $k=2\pi n/L$ with $n=1,2...$.
The kinetic term is
\bea
L_{env}&=&\int dx \: {1\over 2} \partial_{\mu}
\f\:\partial ^\mu \f , \nen
&=&{1\over 2}\sum_{\s=+,-}\sum_k[(\ydot_k^\s)^2 -k^2
( y _k^\s)^2  ] ,
\label{lenv}
\eea
where dots denote time derivative. The system is formed by a
detector that is linearly and locally sensitive to the matter
field with an interacting Lagrangian density
\be
\calL_{int} (x) = -\ve Q\f(x,t) \d(x-X(t)),
\ee
 where $\ve$ is a coupling constant while  $Q(t)$ is the
DeWitt monopole moment\cite{DeW} and  plays the role of $x(t)$
in Section~\ref{intro}. $X(t)$ is the position of the detector
which is assumed here to follow   a classical, definite trajectory
in time.
Performing the trivial spatial integration gives
\bea
 L_{int}   &=& -\ve Q \f(X(t),t )  ,\nen
&=& -\ve Q \sqrt{2\over L} \sum_k[y_k^+cos\: kX(t)
+y_k^-sin\: kX(t)] ,\nen
&\equiv &-\sum_{k \s}Qc_k^\s(t)y_k^\s,
\label{lint}
\eea
where
\bea
c_k^+(t)&=&\ve\sqrt{ 2\over L}\cos  kX(t) ,\nen
c_k^-(t)&=&\ve\sqrt{ 2\over L}\sin kX(t) .
\eea
In this way,     the IF for the observed dipole degree of
freedom $Q$   is computed  from a set of forced,   non-relativistic
harmonic oscillators with a time dependent coupling constant whose
evolution  depends on the trajectory of the detector.
But as it is well known, the Feynman propagator for that case may
be computed exactly\cite{Fey2}. As a result,   the IF (\ref{if2})
can also be obtained and one  gets

 \bea
\G_R[Q_1,Q_2]&=&-\int_0^t ds_1\int_0^{s_1} ds_2
[Q_1(s_1) -Q_2(s_1) ]\m(s_1,s_2) [Q_1(s_2) +Q_2(s_2) ]  ,\nen
\G_I[Q_1,Q_2]&=&\int_0^t ds_1\int_0^{s_1} ds_2  [Q_1(s_1) -Q_2(s_1)]
\n(s_1,s_2) [Q_1(s_2) - Q_2(s_2) ] ,\nen
\label{tem}
\eea
where $\m(s_1,s_2)$ and $\n(s_1,s_2)$ are respectively the noise
and dissipation kernels and depend on $c_k^\s(t)$.
If the detector is following a   trajectory with uniform
acceleration $a$:
\be X(\t)=\inv{a}\cosh(a\t),\ \ \ \ \ \ \ \
t(\t)=\inv{a}\sinh(a\t),\ee
where $\t$ is the proper time,
then in that case one can show that\cite{Hu1}
 \bea
\m(\t_1,\t_2)&=&-\int_0^\infty\ d\o \ I(\o)   \sin{\o(\t_1-\t_2)},
\nen
\n(\t_1,\t_2)&=&\int_0^\infty\ d\o \ I(\o) \coth{ {{\pi\o}
\over a} } \cos{\o(\t_1-\t_2)}
\label{kernel}
\eea
where $I(\o)={{\ve^2}\over {2\pi \o}}$
is the spectral density.
But   except for being written in proper time, the  kernels
(\ref{kernel}) are just those obtained  when the environment
is taken to be a bath of harmonic oscillator at temperature
\be k_BT=a/2\pi \label{Ta} \ee
linearly coupled via an interaction  term of the form
$L_{int}= -\ve \sum_{k \s}Qy_k^\s$ \cite{Fey1,CL}.
In this way, the heating of the vacuum as sensed by an accelerating
detector   is apparent in the IF   formalism
\footnote{One can also define the  interacting Lagrangian to
incorporate an integral over proper time and a delta function over
inertial time (see Ref.\cite{Anglin}). In that setup, the  time
integrals in (\ref{tem}) will come out in proper time. In view
of next section where the proper time will not be unique, we remain
here in inertial time.}.

\section{}
\label{fluc}
\forcepar
In displaying the Unruh effect in the last section, it was assumed
that   the position degree of freedom of the detector was not quantum
mechanical, as the detector followed a well defined, classical
trajectory.  Although this was sufficient to extract the
basic physics of the heating of the vacuum, we know that this
classicality is
an approximation. From a path integral point of view, this
treatment would require that the uniformly accelerating trajectory
sufficiently decohere from other neighboring paths to form a
consistent history~
\footnote{Although sufficient deoherence is usually not the only
criteria for classicality, it is nevertheless essential, and will be
our focus here (see for example Refs. \cite{Ha2,PS}).}.
 But of course the decoherence is always finite at a given time.
We now consider the uncertainty associated with    the fluctuations
in the detector's paths  around the uniformly accelerating trajectory.

For simplicity, we will take the limit where the value of the monopole
degree of freedom $Q$ is given by its quantum mechanical
average. In view of the approximately stationary nature of the
acceleration, this average will also be  assumed to be constant
in time.
As stated in Section~\ref{intro}, our criteria for decoherence will
be $\G_I$. This quantity involves the specification of two paths.
To study the extent to which the acceleration is well defined, we
take two uniformly accelerating trajectories with slightly different
accelerations:

\be
x_i=\sqrt{a_i^{-2}+t^2}
 \ee
for $i=1,2$.  After defining,
\bea
g_{1k}^+&\equiv& \ve Q\sqrt{ 2\over L}\cos  k\sqrt{a_1^{-2}+t^2} ,\nen
g_{1k}^-&\equiv& \ve Q\sqrt{ 2\over L}\sin  k\sqrt{a_1^{-2}+t^2} ,\nen
g_{2k}^+&\equiv& \ve Q\sqrt{ 2\over L}\cos  k\sqrt{a_2^{-2}+t^2} ,\nen
g_{2k}^-&\equiv& \ve Q\sqrt{ 2\over L}\sin  k\sqrt{a_2^{-2}+t^2} ,
\label{g}
\eea
we have  for  the Lagrangians (\ref{lenv}) and  (\ref{lint}) the
result\cite{Hu1}:
\bea
\G_I(a_1,a_2,t)&=&{1\over 2}\sum_{\s=+,-}\sum_k\int_0^t
ds_1\int_0^{s_1}
\frac{ds_2}{k}  [  g_{1k}^\s(s_1) - g_{2k}^\s(s_1) ]\nen
&&\hspace{1.9 in}\cos{ (s_1-s_2)} [ g_{1k}^\s(s_2) -
g_{ 2k}^\s(s_2) ] ,\nen
&& \nen
&=&\frac{\ve^2 Q^2}{2\pi}\int_0^t ds_1\int_0^{s_1}
ds_2 \int_0^\infty
  \frac{dk}{k}  [\cos kA+\cos kB
\nen
&&\hspace{1.9 in}
-\cos kC-\cos kD]\cos kF ,\nen
&=&\frac{\ve^2 Q^2}{8\pi}\int_0^t ds_1\int_0^{s_1} ds_2\
P(s_1,s_2,a_1,a_2).
\label{def}
\eea
In the second equality of (\ref{def}), we have done the sum
over $\s$, defined the quantities
\bea
A&\equiv&\sqrt{a_1^{-2}+s_1^2}-\sqrt{a_1^{-2}+s_2^2},\nen
B&\equiv&\sqrt{a_2^{-2}+s_1^2}-\sqrt{a_2^{-2}+s_2^2},\nen
C&\equiv&\sqrt{a_1^{-2}+s_1^2}-\sqrt{a_2^{-2}+s_2^2},\nen
D&\equiv&\sqrt{a_2^{-2}+s_1^2}-\sqrt{a_1^{-2}+s_2^2}, \nen
F&\equiv&s_1-s_2,
\eea
and took the continuum limit for $k$ by substituting
$\sum_k\rightarrow\frac{L}{2\pi}\int_0^\infty dk$. In the
last equality  of (\ref{def}), the $k$ integration was performed
by making use of\cite{GR}
\be
\int_0^\infty dx \frac{\sin ax \sin bx}{x}=\frac{ 1}{2}\ln{
\left| \frac{a+b}{a-b}\right|}
\ee
and we also  defined
\be
 P(s_1,s_2,a_1,a_2)\equiv \ln \left(
 \frac{C^2-F^2}{A^2-F^2 }\ \ \frac{ D^2-F^2}{B^2-F^2 }
 \right)^2.
\label{P}
\ee
Note that in (\ref{P}), as $s_2\rightarrow s_1$,  $A^2-F^2$ and
$B^2-F^2$ vanish and $P$ is divergent on that ``line''. Depending
on the values of $s_1,a_1,a_2$, there will generally also be
values of $s_2$ for which $C^2-F^2$ or $D^2-F^2$ are vanishing.
But since those divergencies are logarithmic, the decoherence
remains nevertheless finite, even without imposition of  a
ultra-violet  cut-off.
This is in contrast with what is usually   done in the
Caldeira-Legget model\cite{CL} and in the minisuperspace
approximation of quantum cosmology (see for example \cite{PS}).

We now wish to extract the  features of $\G_I$ for paths
lasting  a time large compared with the inverse accelerations
($t a_1\gg 1,
t a_2\gg 1$). Consider the rate of increase of $\G_I$ with time.
{}From (\ref{def}),
\be
{\dot{\G}}_I (t,a_1,a_2)=\frac{\ve^2 Q^2}{8\pi} t \ R(t,a_m,a_d),
\label{bu}
\ee
with:
\bea
R(t,a_m,a_d)&\equiv&
\int_0^1dy \
P(s_1=t,s_2=yt,a_1,a_2),\nen
&=& R(\tilde{t}, \tilde{a}_d) .
\label{bi}
\eea
In (\ref{bu}), we made the change of variable $s_2=yt$ ($y$
is unitless), and worked with the average $a_m\equiv
\inv{2}(a_1^{-1}+a_2^{-1})$ and difference $a_d\equiv
(a_1^{-1}-a_2^{-1})$ in {\it inverse}  accelerations.
In the last line of (\ref{bi}), $a_m$ was used for the
scale factor in $P$, namely:  $\tilde{t}\equiv t/a_m$ and
$\tilde{a_d}\equiv a_d/a_m$.
Now except for a transient time $\tt{\ \lower-1.2pt\vbox{\hbox{\rlap{$<$}
\lower5pt\vbox{\hbox{$\sim$}}}}\ }1$,
we have
\be
R(\tilde{t}, \tilde{a}_d)\ \rightarrow \ f(\tilde{a}_d),
\ee
that is, $R$ is time independent  to a good approximation.
(Obviously, $f(\tilde{a}_d)$ is an even function.)
This can be seen in Fig.1, where numerical evaluations of $R$
are given  as a function of $\tt$ for a few representative values
of $\adt$. Note that for $\tt<1$, $R$ takes large values, and  in
fact diverge as $\tt\rightarrow 0$, since in (\ref{P}), $A^2-F^2$
and $B^2-F^2$ vanish in that limit. But when multiplied by $t$
in (\ref{bu}),
${\dot{\G}}_I$ remains obviously  finite. For example, at
$\adt=0.1$, the peak in $tR$ occurs for $\tt\approx
 \adt$ and has height 0.7. As $\tt$ grows beyond unity, the time
development of ${\dot{\G}}_I$ is clearly linear to a very good
approximation. In the computation of ${\dot{\G}}_I$, the
importance of the initial peak becomes negligible in the
limit of small $\adt$ and large $\tt$. In that regime, (\ref{bu})
integrates to
\be
 {\G}_I(t,a_1,a_2)\approx\frac{\ve^2 Q^2}{16\pi} t^2 f(\adt).
  \label{bi2}
\ee

Using (\ref{bi}), (with (\ref{P})), a simple and rough estimate can
be obtained for $f(\adt)$ by the substitutions
\bea
\sqrt{(1\pm \inv{2} \adt)^2+\tt^2}\ &\rightarrow &\  \tt,\nen
\sqrt{(1\pm \inv{2} \adt)^2+\tt^2y^2}\ &\rightarrow &\  \tt y,
\label{sqrt}
\eea
{\it after} multiplying out the terms in (\ref{P}) (e.g. $C^2-F^2
\rightarrow  2a_m^2+\inv{2}a_d^2$). Eq. (\ref{sqrt}) is clearly a
poor substitution around $y=0$ and   also fails to  reproduce any
of the divergencies present in $P$ mentioned above. To leading order
in $\adt$, it leads to $f(\adt)=2 \adt^2$. The actual decoherence is
in fact larger than  what this predicts. But  as   $\adt$ gets smaller,
it is in fact quite good, up to a multiplicative constant $N$.
The Fig.2 presents  the exact, numerically computed  result for
$f(\adt)$ in the range
$10^{-4}<\adt < 10^{-1}$ (obtained with $\tt=10$). It also includes
the curve  $N \adt^2$ for $N=10$ and $N=110$. It is seen that $f(\adt)$
approaches a quadratic behavior as $\adt$ decreases. With (\ref{bi2})
we thus have for $|\adt |\ll 1$ and $\tt \gg 1$ the approximation
\be
{\G}_I(t,a_1,a_2)\approx\frac{N}{16\pi} \ve^2 \langle Q\rangle^2
 t^2 \left ( \frac{a_1-a_2}{(a_1+a_2)/2}\right ) ^2,
\label{deco}
\ee
where $Q$ was replaced by its quantum mechanical average (quantum
fluctuations of $Q$ are neglected).

We now take the condition $\G_I \approx 1$ as  separating the decohering
and non-decohering accelerations.
Using (\ref{deco}) and (\ref{Ta}), we conclude that a detector on a
decohered accelerating trajectory for a time $t$ will  be subject to
thermal fluctuations of the order:
\be
\left | \frac{\D T}{T} \right |\approx \inv{|\ve | \langle Q\rangle t}
\label{delta}
\ee
where we dropped a constant of order one.  Eq.  (\ref{delta})  can
also be  understood as an uncertainty relation between the `time of
flight' of the detector  and the spread in temperature, with the
uncertainty given by $\inv{|\ve | \langle Q\rangle}$ and valid for
$t/T > 1$.

\section{} \label{conc}
In Eq. (\ref{deco}), the quadratic dependance of the decoherence on
the coupling to the bath was already apparent from (\ref{g}). It is
general for a system coupled linearly to the bath
(by construction, the
amplitude of the  detector's excitations act to modify  the
effective coupling).
It is interesting to see that for the class of paths considered here,
the dependance  in time is also quadratic.
This shows that   the time is also making the coupling constant stronger
or inversely, that a larger coupling constant is identical to a longer
time evolution.
Qualitatively, this is in line with the result of Ref. \cite{CM}. As the
coupling to the bath gets stronger and that one waits longer, clearly
more particles are created and decoherence is
increased.
This can also be seen through (\ref{if2}). When particles are copiously
produced, a slight difference in the two paths will easily make  the
two states orthogonal.

Issues of decoherence similar to the ones considered here  could well
prove to be relevant in the
context of black holes. As it is well known,  one of the  distinguishing
features   of black holes is their  ability to Hawking radiate, in close
connection with the uniformly accelerating detector.  The environment
given  by these radiated particles could  then  serve in the  study
of the validity of the semi-classical approximation\cite{PS,sca}.
\vskip 0.5in
\begin{centerline}
{\bf ACKNOWLEDGMENTS}
\end{centerline}
\noindent
I am grateful to  Samir Mathur for a number of discussions in relation
with this work.

\newpage
\noindent
{\bf
\large{{\bf Figure Captions}} }\\  \\
Fig.1 : Plot of $R(\tt,\adt)$ as a function of the reduced time
$\tt$, for $\adt=0.05$ (curve {\bf A}), $\adt=0.075$ (curve {\bf B})
and $\adt=0.1$ (curve {\bf C}). It is apparent  that (for a
given $\adt$) as  $\tt{\ \lower-1.2pt\vbox{\hbox{\rlap{$>$}\lower5pt
\vbox{\hbox{$\sim$}}}}\ }1$, $R(\tt,\adt)$ is well approximated by
a constant.
\\  \\
Fig.2 : Logarithmic plot of $f(\adt)$  (curve {\bf A}) for a range of
$\adt$. The function $N \adt^2$ is also presented for $N=110$
(curve {\bf B}) and $N=10$ (curve {\bf C}). As
$\adt$ decreases, the behavior of $f(\adt)$ becomes nearly
quadratic.


\newpage
\begin{thebibliography}{99}
\bibitem{Fey1}{R. Feynman and F. L. Vernon,  \ap{24} (1963) 118.}
\bibitem{ZCYC}{Z. Su, L.-Y. Chen, X. Yu and K. Chou,  \prb{37} (1988)
9810, and references therein; E. Calzetta and B.L. Hu,
\prd{35} 495 (1987). On the Close-Time-Path formalism, see the
reviews: N. P. Landsman and Ch. van Weert, \prep{145} (1987) 141 ;
K. Chou, Z. Su, B. Hao and L. Yu, \prep{118} (1985) 1. }

\bibitem{Zu}{W. H. Zurek,  \prd{24} 1516 (1981); \prd{26} 186
(1982); \pto{44} (1991) 36; M. Gell-Mann and J. B. Hartle, in
{\it Complexity, Entropy and the Physics of Information}, edited
by W. H. Zurek, Santa Fe Institute Studies in the Sciences of
Complexity Vol. VIII (Addison-Wesley, Reading, MA,  1990).}

\bibitem{Hal}{H. F. Dowker and  J. J. Halliwell,  \prd{46}
(1992) 1580.}

\bibitem{Hu1} {B. L. Hu and  A. Matacz, Maryland U. preprint
UMDPP-93-210.}

\bibitem{Un}{W. G. Unruh,  \prd{14} (1976) 870.}

\bibitem{Anglin}{J. R. Anglin, \prd{47} (1993) 4525.}


\bibitem{Har}{J. B. Hartle, in {\it Quantum Cosmology and
Baby Universes},
Proceedings of the 7th Jerusalem Winter School,
edited by S. Coleman, J. Hartle, T. Piran and S.
Weinberg, Jerusalem, Israel, 1990, (World Scientific,
Singapore, 1991).}

\bibitem{DeW}{B. S. DeWitt, in   {\it General Relativity: An
Einstein Centennial Survey}, edited by S. W. Hawking and
W. Israel (Cambridge University Press, Cambridge, England,
1979), p. 680.}

\bibitem{Fey2}{R. Feynman and C. R. Hibbs, {\it Quantum
Mechanics and Path Integrals}, (McGraw-Hill, New-York, 1965)}

\bibitem{CL}{A. O. Caldeira and A. J. Legget, \pa{121} (1983) 587.}

\bibitem{Ha2}{J. J. Halliwell,  \prd{48} (1993) 4785.}

\bibitem{GR}{I. S. Gradshteyn and I. M. Ryzhik, {\it Tables of
Integrals, Series, and Products}, 4th ed. (Academic, San Diego,
CA, 1980), p. 414, Eq. 3-741(1).}
\bibitem{CM}{E. Calzetta and D. Mazzitelli, \prd{42} (1990) 4066.}

\bibitem{PS}{ J.P. Paz and  S. Sinha, \prd{44} (1991) 1038;
\prd{45} (1992) 2823 }


\bibitem{sca}{See for example J. B. Hartle, in {\it
Gravitation in Astrophysics} (Cargese 1986), Proceedings
of the Summer Institute, Cargese, France, 1986, edited by
B. Carter and J. Hartle, NATO ASI Series B: Physics, Vol.
156 (Plenum, New York, 1987);
T. Padmanabhan,
\cqg{6} (1989) 533; T. Padmanabhan and  T.P. Singh, \cqg{7}
(1990) 411; S. D. Mathur, MIT-CTP-2304 (hep/th 9404135);
E. Keski-Vakkuri, G. Lifschytz, S. D. Mathur and M. Ortiz,
MIT-CTP-2341.}


\end{thebibliography}
\end{document}